%% file: v23windows_paul.tex
\documentclass{v23windows}

\usepackage[nottoc]{tocbibind}  %
\usepackage{cite}               %
\usepackage{caption}            %
\usepackage{multirow}           %
\usepackage{packages}
\usepackage{nccmath}
\usepackage{authblk}
\makeatletter
\@ifpackageloaded{tikz}%
{}%
{\usepackage{tikz}}%
\makeatother

\usepackage{setspace}
\AtBeginDocument{%
  \addtolength\abovedisplayskip{-0.5\baselineskip}%
  \addtolength\belowdisplayskip{-0.5\baselineskip}%
  }
  
  \hypersetup{%
  pdfborder={0 0 0.1}  %
}

\graphicspath{%
  {./},%
  {images/},%
 }

\usepackage{macros}
\usepackage{here}
\usepackage{subfig} 
\newlength\textwidthcm
\newlength{\textwidthPlot}
\newlength{\totalPlotWidth}
\newlength{\twoPlotWidth}
\newlength{\twoPlotSpacing}
\newlength{\threePlotWidth}
\newlength{\threePlotSpacing}
\newlength{\threePlotSmallWidth}
\newlength{\threePlotSmallSpacing}
\newlength{\fourPlotWidth}
\newlength{\fourPlotSpacing}
\newcommand{\Photo}{\includegraphics[height=35mm]{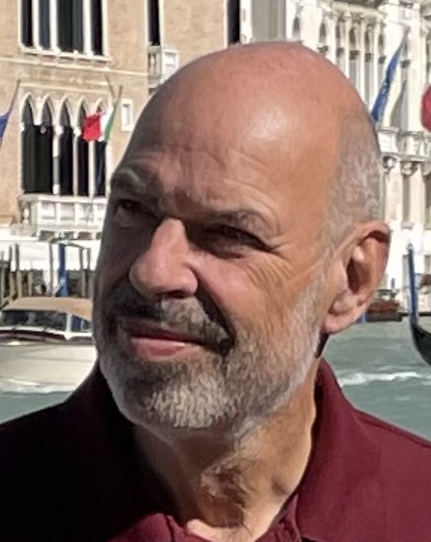}}

\begin{document}
\vspace*{4cm}

\title{Measuring the Electric and Magnetic Form Factors of Protons and Antiprotons\\ at small \QTwo and the charge radii of hadrons}

\author{Stephan Paul}

\address{Stephan Paul\\member of the AMBER collaboration\\Technical University Munich and Max-Planck-Institute for Physics, Germany}

\maketitle\abstracts{
Charge radii of hadrons are back in the focus since precision spectroscopic measurements in muonic hydrogen revealed a proton charge radius considerably smaller than 
generally accepted ever since R.~Hofstaedter's pioneering experiments. Recent experimental results also point to an underestimation of systematic uncertainties in most previous measurements. A new method was thus proposed by AMBER at CERN using elastic scattering by very high energy muons using an active hydrogen target. With a simplified setup we may resume elastic scattering of $\pi$ and K mesons on electrons in inverse kinematics, and, for the first time, probe antiprotons with few \% precision for their charge radius. Applying this method also for protons allows a separation of \GEP and \GMP down to \QTwo=\SI{1 e-3}{\GeVcc}.
}

\input{Proceedings-Vietnam-2023}



\bibliographystyle{utphys_bgrube}
\bibliography{form_factor.bib}



\end{document}

%% file: Proceedings-Vietnam-2023.tex
\section*{Introduction}
\label{sec:Introduction}
The structure of hadrons, their charge distribution and their moments are key to understanding strongly interacting systems.
For baryons, these quantities are encoded in the electric and magnetic form factors, but also in their polarizabilities, measurable quantities that depend on the four momentum transfer squared (\QTwo) as a single kinematic variable corresponding to the resolution at which they are probed. The electric and magnetic form factors \GE and \GM can be interpreted as Fourier transforms of the electric charge and current distributions using the Breit frame. Their dependence on \QTwo is a priori unknown, making their extrapolation into unmeasured regions very questionable. In recent years this has led to much discussed results on the extraction of the proton charge radius, defined as the slope of \GEP at \QTwo$=0$. Alternatively, charge radii can be derived from a shift in the atomic S-states in hydrogen, as these are particularly sensitive to the charge distribution at the origin. Muonic hydrogen is particularly useful since the Bohr radius scales inversely with the reduced mass, i.e. the mass of the bound lepton. The variety of measurements made over the last 10 years for \RP gives a very inconclusive picture, with the muonic hydrogen data~\refCite{Pohl:2014bpa} standing out for its unique precision (see \cref{fig::PR_status}). Recent experimental results indicate an underestimation of the systematic uncertainties for most previous measurements. The situation is exemplified for elastic electron scattering in \cref{fig::AMBER_PR}, which shows data from~\refCite{A1:2013fsc,xiong2019small} that are apparently incompatible and lead to a different value for \RP. Many new measurements are in preparation. Much effort has recently been made to extend the measurements of \GEP to even smaller values of \QTwo, although there is little direct sensitivity to the nucleon charge radius~\refCite{Suda::2018,xiong2019small}. In addition, radiative effects are large at small \QTwo, requiring corrections that are often much larger than experimental acceptance effects or other measurement systematics. 

Meson charge radii have recently become more relevant, as octet mesons play a key role in the production of hadron mass and, as Goldstone bosons, in spontaneous chiral symmetry breaking. Experimentally, they have been studied in inverse kinematics experiments. Modern high-rate experiments can dramatically improve the precision of these measurements. The measurement of charge radii in inverse kinematics requires extracted beams at high energy, which have largely disappeared.  AMBER is in the unique position of being a high resolution, high rate spectrometer with secondary and tertiary beams available in a wide energy range of \SIrange{50}{280}{GeV}. 

\section*{Measurement of \GEP in elastic $\mu -p$ scattering}
\label{sec:mup}
\begin{figure}[htbp]
\begin{center}
        \subfloat[][]{%
        \label{fig::PR_status}
        {\includegraphics[width=0.95\twoPlotWidth]{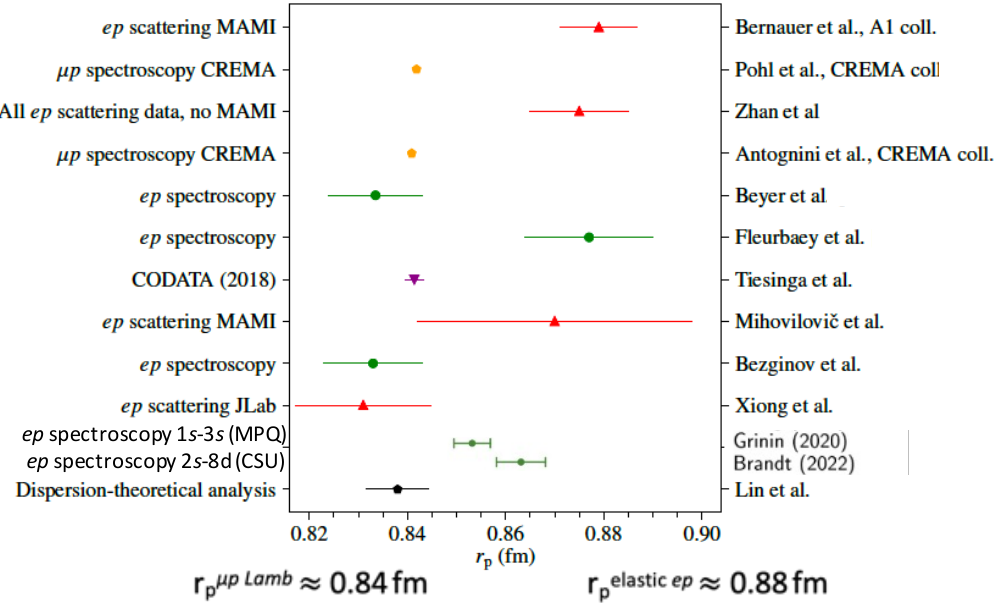}}}
        \hspace*{\twoPlotSpacing}%
        \subfloat[][]{%
        \label{fig::AMBER_PR}
        {\includegraphics[width=0.92\twoPlotWidth]{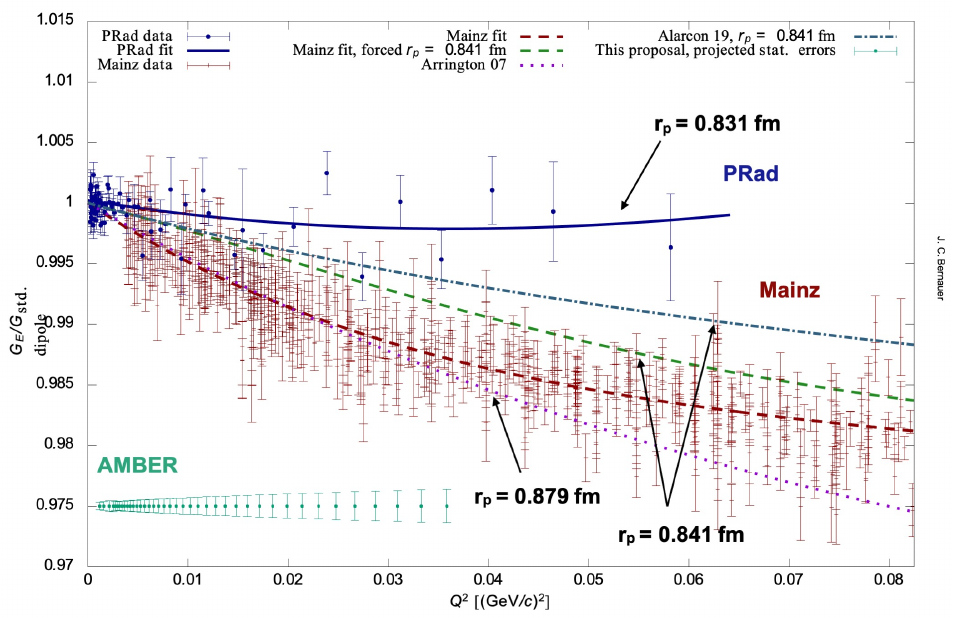}}}
        \caption{Left: Summary of experimental results for \RP. Right: \GEP for the Mainz~\refCite{A1:2013fsc} and PRAD~\refCite{xiong2019small} experiments. The green dots indicate the expected statistical uncertainties from this measurement~\refCite{AMBER_Proposal:2019}.}
\end{center}       
\end{figure}

Recently, a new method has been proposed to measure \GEP using elastic scattering of very high energy muons on target protons and detection of the recoil proton within an active target with high precision~\refCite{AMBER_Proposal:2019}. This method benefits from radiation effects below 1\% (lepton bremsstrahlung) and can reconstruct the kinematics in several ways. Due to the high energy of the incoming muon and the rather low values of the addressable \QTwo, this experiment offers low sensitivity for \GMP.  The experimental resolution allows a minimum \QTwo of \SI{e-3}{\GeVcsq}, the maximum is \SI{2e-2}{\GeVcsq} if the recoil proton is detected, much larger if we rely on the scattered muon. The measurements will use the AMBER spectrometer at the CERN SPS with a dedicated target arrangement. The central element is an active target consisting of a hydrogen-filled TPC operated at a high pressure of \SIrange{4}{20}{bar} and surrounded by low-mass tracking detectors consisting of \SI{500}{\micro\meter} thick scintillating fibers and pixel detectors. The recoil proton emitted at a large angle \wrt the incoming beam particle is stopped in the TPC and its total energy is measured with \SI{70}{keV} resolution and the muon scattering angle is measured to better than \SI{25}{\MuRadian}. There will thus be two independent measurements \QTwo, which will allow most of the background reactions to be removed. The data planned for 2025 should allow an overall precision of about 1\% for \RP (see \cref{fig::AMBER_PR}).

\vspace{-0.5cm}
\section*{Measurement of \GEP and \GMP in elastic $\textrm{p-e}$ scattering}
An alternative to uncertain extrapolations of \GEP to \QTwo$ = 0$ is the use of an internally consistent framework that uses all available data on nucleon form factors, both in the time and space domain, and for both isospin states, neutron and proton. Dispersion relations relate the form factors to the hadronic mass spectrum, thereby gaining stability in the extraction of \RP~\refCite{Lin:2021umz}.  The accuracy of such an analysis is limited by, among other things, the lack of information on \GMP at small \QTwo.
The most comprehensive measurement of the proton form factors to date has been carried out by the Mainz group~\refCite{A1:2013fsc} using low-energy electron scattering. They performed a Rosenbluth separation of \GEP and \GMP within a \QTwo range of \SIrange{0.0152}{0.552}{\GeVcsq}. A new experiment is in preparation to extend the available \QTwo range for \GEP by elastic scattering of very low energy electrons in the range of \SIrange{20}{100}{\MeV}.~\refCite{Suda::2018}. Such measurements suffer from poor resolution due to multiple scattering, which limits angular and momentum measurements.
\par
The measurement described above
offers little sensitivity to \GMP because the incoming muon is of very high energy und thus the longitudinal photon polarization $\epsilon$ = 1. Here, we present a novel scheme that combines the advantage of using high energy particles with sensitivity to both \GEP and \GMP. Using elastic proton-electron scattering at energies of tens to hundreds of GeV in inverse scattering kinematics, we can detect both particles emerging from the elastic scattering process. This method was developed about 40 years ago to measure the electric charge radius of unstable particles such as pions, kaons and $\Sigma$-hyperons. While the drawback of inverse kinematics is the limited upper range in \QTwo due to the unfavorable mass difference between the projectile (heavy) and the target (light), we can now address the form factors down to very small values of \QTwo, although previous experimental setups have never been able to exploit this virtue.

This paper presents the first ideas for applying the inverse scattering kinematics scheme to measure \GEP and \GMP over a large range of \QTwo, and to probe the charge distribution of antiprotons for the first time. In inverse kinematics we scatter on quasi-free atomic electrons, which allows the use of solid targets (e.g. Be). This allows very high luminosities compared to a gaseous TPC, as discussed above.  We can use three kinematic observables to determine \QTwo whose relative weights depend on \QTwo, namely $\theta^{\textrm{scatt}}_p$, $\theta^{\textrm{recoil}}_e$ and $\textrm{p}^{\textrm{scatt}}_p$. The relative \QTwo resolution should be kept below 10\%. It depends on the multiple scattering in the target material and the angular resolution of the reconstructed tracks.  The target thickness must be small (\SI{1}{mm}) for the lowest achievable \QTwo where electron energies are very low, and can increase to \SI{2.5}{cm} for high \QTwo. 

It is the virtue of inverse kinematics that we can vary the photon polarization $\epsilon$ for a given value of \QTwo using different beam energies within the \SIrange{50}{280}{GeV}. The reason for this is the strong dependence of the maximum achievable value $\textrm{Q}_{\textrm{max}}$ of \QTwo on the beam energy (and on the beam particle mass) and the strong decrease of the electron scattering angle towards $Q_{max}$ (see \cref{fig::GE_proton}). 
We can measure the differential cross section at different beam energies 
and target thickness (thin targets for high resolutions at very small \QTwo, thick targets to reduce statistical uncertainties at high \QTwo). 
We can thus perform the classical Rosenbluth separation over a wide range of \SI{1 e-4}{\GeVcsq} < \QTwo < \SI{4 e-2}{\GeVcsq} and separate electric and magnetic form factors (see left panel of \cref{fig::GE_proton}). The lower limit for such separation is determined by the lowest possible beam energy and the multiple scattering in the target (count rates are not an issue at very low values of \QTwo). As the contribution of the magnetic form factor is weighted by \QTwo, this technique is effectively limited to \QTwo=\SI{e-3}{\GeVcsq}, where we could achieve a statistical uncertainty for \GMP of about 20\% within a bin of \SI{e-3}{\GeVcsq}. This measurement would allow a statistical accuracy for the extraction of \RP well below 1\%, but many data sets taken at different beam energies and target thicknesses will have to be combined, introducing systematic uncertainties (normalization, efficiencies) absent for $\mu-\textrm{p}$ scattering discussed above.

\section*{The charge radius of antiprotons}
\label{sec:Antiprotons}

\begin{figure}[htbp]
\begin{center}
        \subfloat[][]{%
        \label{fig::GE_proton}
        {\includegraphics[width=0.92\twoPlotWidth]{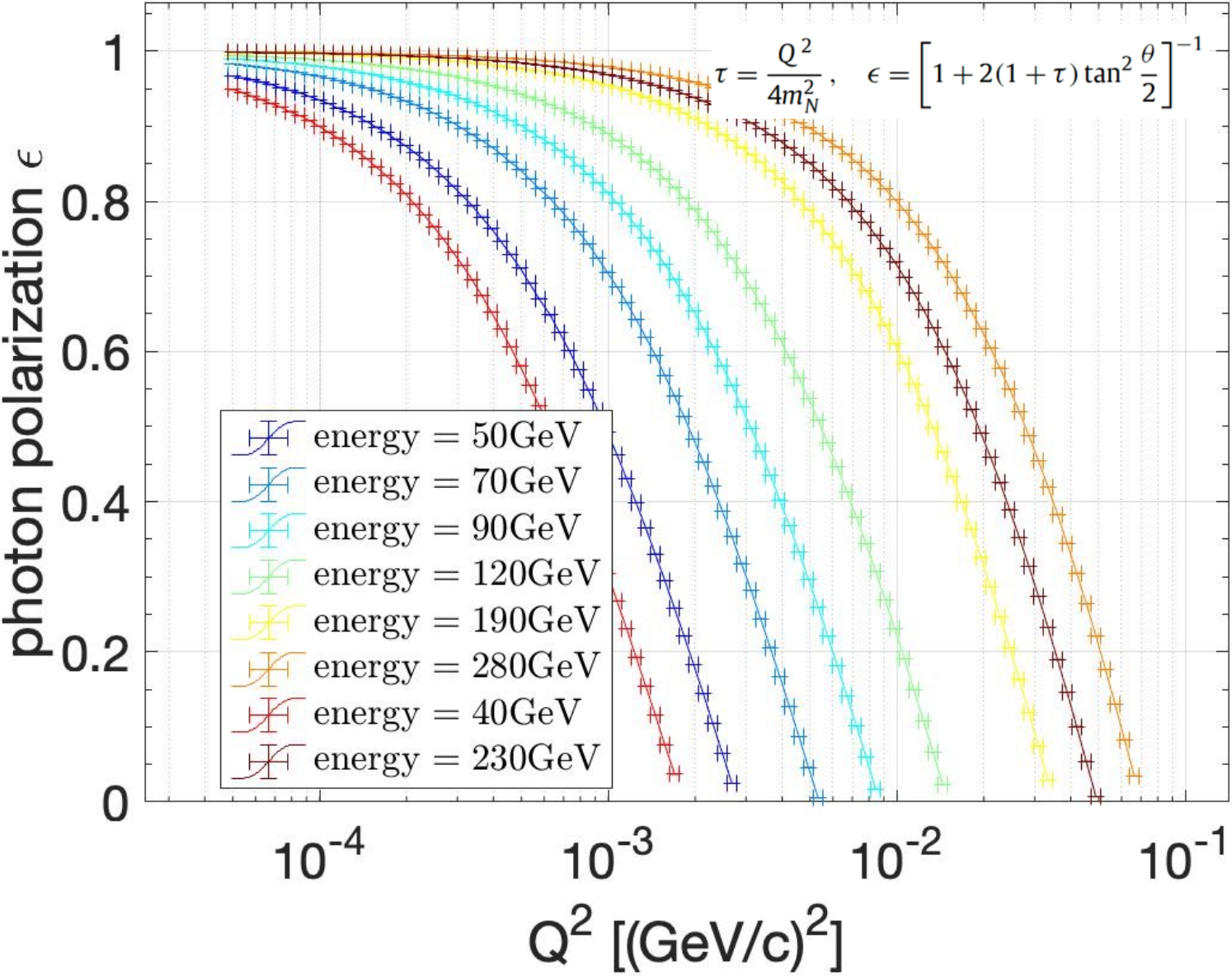}}}
        \hspace*{\twoPlotSpacing}%
        \subfloat[][]{%
        \label{fig::beam_fraction}
        {\includegraphics[width=1.0\twoPlotWidth]{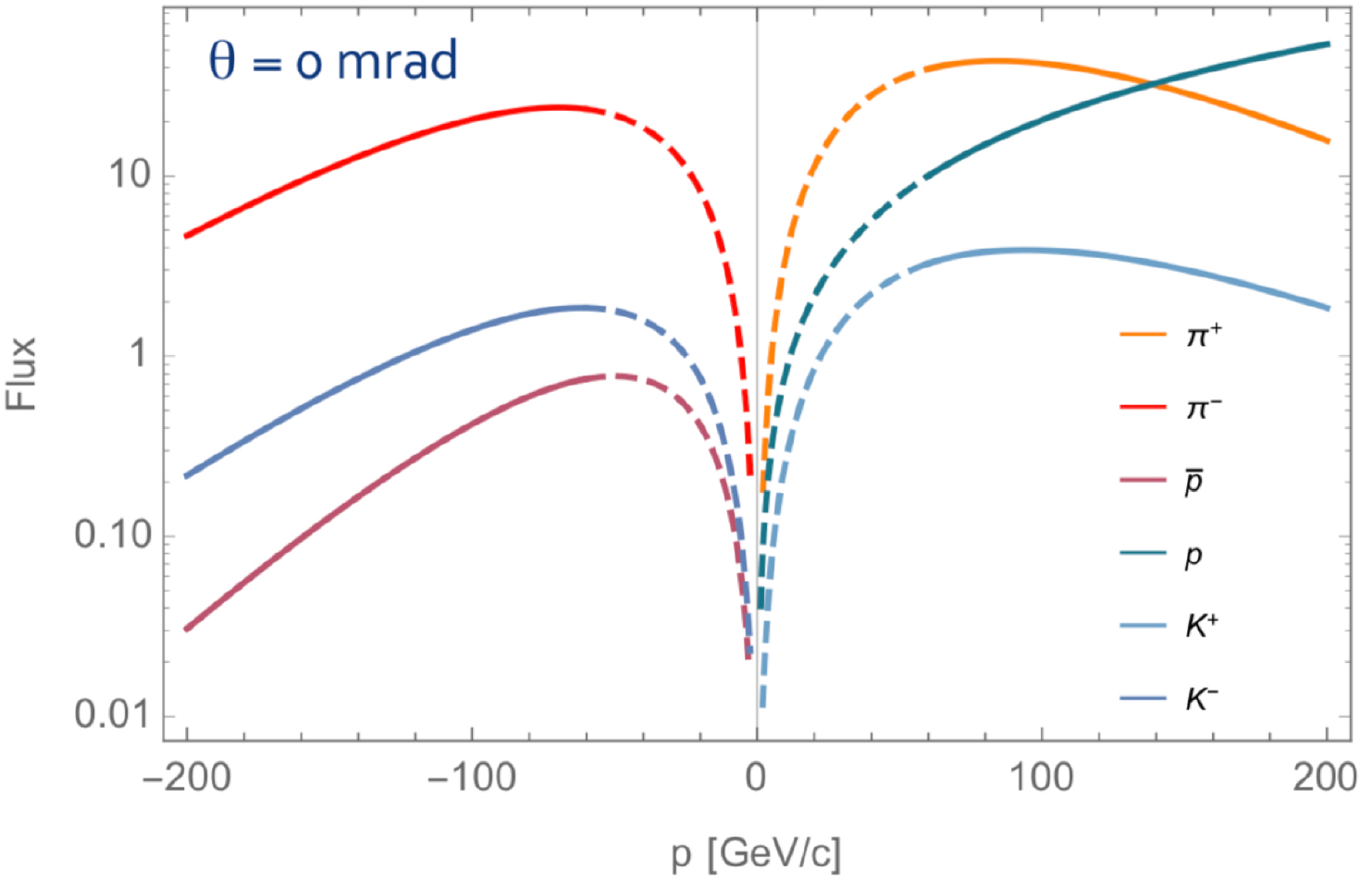}}}
        \caption{Left: Variation of the photon polarisation $\epsilon(\QTwo)$ for different beam energies in inverse kinematics for incoming protons, relevant for a Rosenbluth separation of \GE and \GM. Right: Beam composition at the experiment target on the M2 beam line at the CERN SPS~\refCite{CERN_M2:2022}.}
\end{center}
\end{figure}

The virtue of the M2 beam line at CERN is depicted for hadrons in \cref{fig::beam_fraction}.  For negative beams, the sizable fraction of $K^-$ and \pbar allows to simultaneously perform the same measurement for all three hadron species.
   
      The experiment for \pbar can be performed in the same way as for protons. However, the data at the highest values of \QTwo require special care, as the antiproton fraction drops below 1\% at very high momenta. For the determination of the electric and magnetic charge radii, however, they give the highest sensitivity (see \refCite{Lee:2015jqa} for a detailed discussion). Two issues need to be addressed: a) the background due to beam pions and b) the counting rates. While the latter could be compensated for by using a thick target, the former requires excellent particle identification using two-beam Cherenkov counters (CEDARs), with the residual background from kaons and pions in particular to be further removed by kinematic correlations. While CEDARs have excellent particle identification at lower momenta (e.g. a kaon selection by CEDARs at \SI{190}{\GeVc} leaves about 1\% of misidentified pions) it steeply drips towards the highes momenta. However,  at the highest beam momentum of \SI{280}{\GeVc}, \QMax = \SI{0.07}{\GeVcsq} compared to \QMax = \SI{0.035}{\GeVcsq} at \SI{190}{\GeVc}. 
            
We estimate the precision obtained for the charge radius 
$\sqrt{\RAP}$ of the antiproton and the electric form factor \GEAP analogue to the case of the proton. However, the \pbar flux is about a factor of 50-100 lower. At this stage we expect to run in parallel with the pion and kaon form factor measurements. 

\begin{figure}[htbp]
\begin{center}
        \subfloat[][]{%
        \label{fig::fit_antiproton_ff_pion_running}
        {\includegraphics[width=0.92\threePlotWidth]{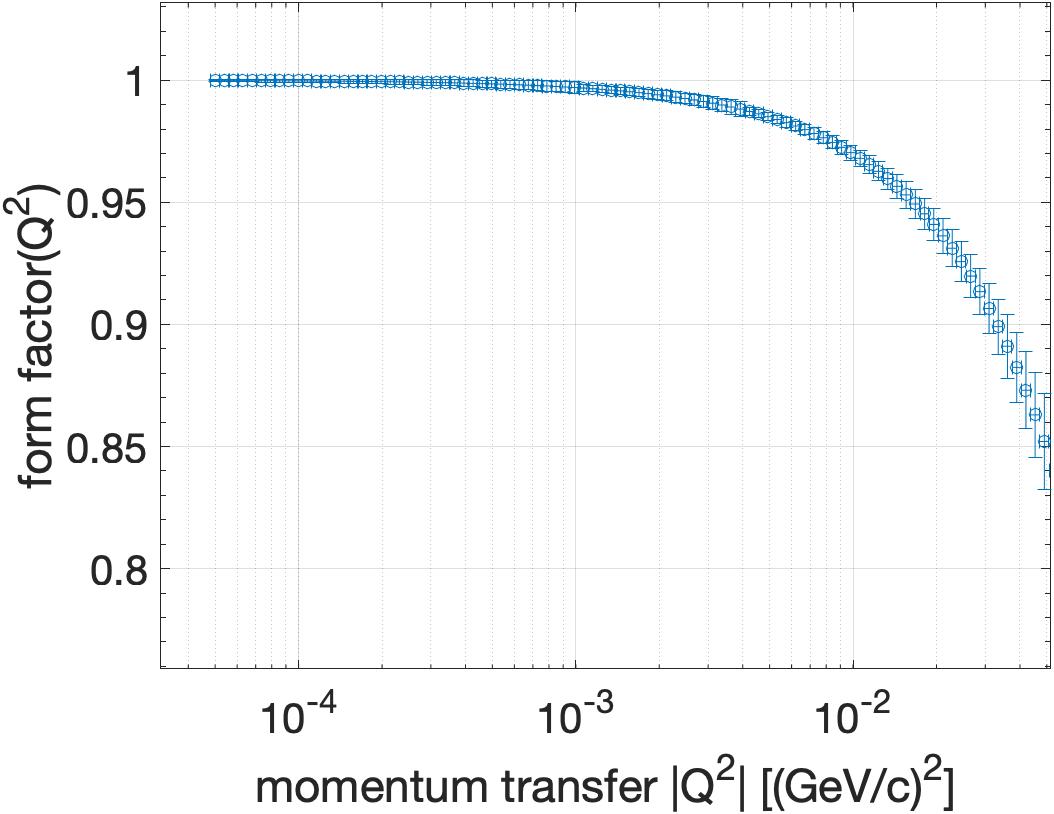}}}
        \hspace*{\threePlotSpacing}%
        \subfloat[][]{%
        \label{fig::kaon_ff}
        {\includegraphics[width=0.92\threePlotWidth]{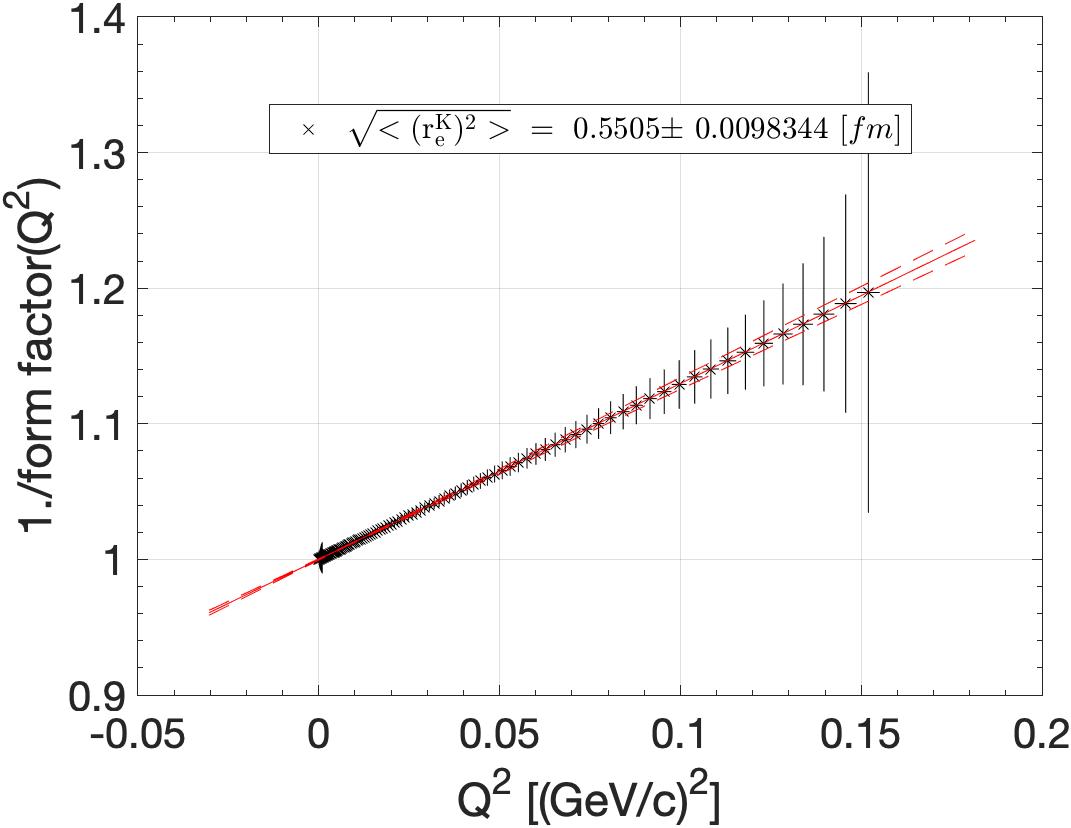}}}
        \hspace*{\threePlotSpacing}%
        \subfloat[][]{%
        \label{fig::pion_ff}
         {\includegraphics[width=0.92\threePlotWidth]{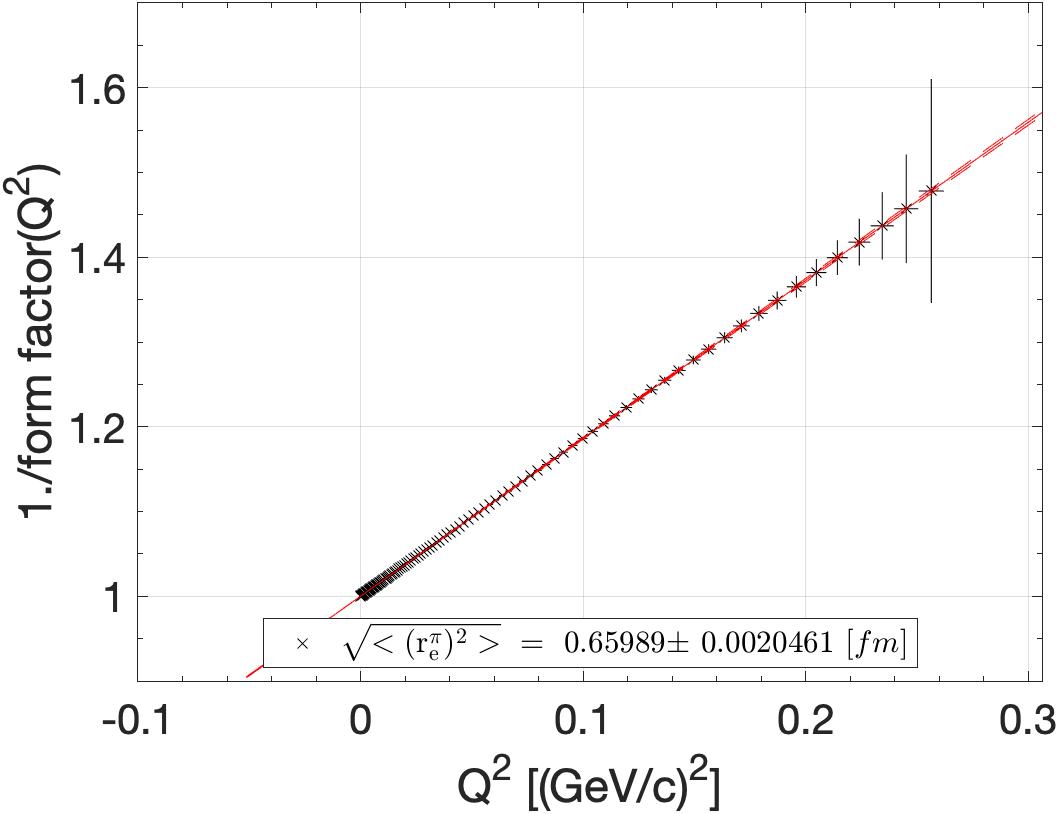}}}                    
\caption{a) Measurement $\sigma_{meas}\left (\QTwo\right )/ \sigma_{point}\left (\QTwo \right )$ for antiprotons in a dedicated run. b-c) : inverse of the meson form factors obtained with four different sets of energy/target thickness combinations together with a fit for meson radii. b) for $K^-$;  c) for $\pi^-$}
\end{center}
\end{figure}

We can also consider a separate running strategy for antiprotons to optimize for a Rosenbluth separation of the form factors. For simplicity, we assume the same running strategy as for the analogue physics program for protons. With a dedicated running strategy as for protons, we can perform the first measurement of \GEAP \cref{fig::fit_antiproton_ff_pion_running}. However, the uncertainties for \GMAP will be large.

\section*{Charge radii for pions and kaons}
\label{sec:mesons}
Several experiments have used inverse kinematics to determine the charge radii of pions and kaons. By fitting a functional dipole form to the differential cross section, accuracies of 2\% for pions \cref{fig::pion_ff}~\refCite{AMENDOLIA1986168} and 6\% for kaons \cref{fig::kaon_ff}~\refCite{Amendolia:1986ui} have been obtained. In parallel to the measurements with antiprotons, we can significantly improve the precision of the differential cross section for $\pi-e$ and $K-e$ scattering and also obtain high-precision measurements for both meson types. We will be helped by the high quality of the reconstruction, the analysis methods and the much higher beam fluxes of modern high-rate experiments. We will also obtain smaller values of \QTwo, which will allow more confident extrapolations. The inverse of the form factors are shown in \cref{fig::kaon_ff} and \cref{fig::pion_ff} together with the results of a linear fit to the charge radii.

\section*{Conclusions}
The location of AMBER on a high intensity secondary/tertiary beam line allows precision experiments on both the baryonic form factors and the charge radii of hadrons. The use of inverse kinematics allows high statistical accuracy, extension to very low values of \QTwo and a Rosenbluth separation of \GM and \GE. For the first time, we can address the structure of antiprotons.